\documentclass{iopart}
\expandafter\let\csname equation*\endcsname\relax
\expandafter\let\csname endequation*\endcsname\relax
\usepackage{euscript,amsmath,amssymb,amsfonts,graphicx,bm}

\usepackage{float}
\usepackage{epsfig}
\usepackage{esint}
\usepackage{chemfig}
\usepackage{cite}

\usepackage{bm,braket}

\eqnobysec

\newcommand{\calM}{\mathcal{M}}

\renewcommand{\l}{\left(}
\renewcommand{\r}{\right)}

\renewcommand{\theequation}{\arabic{section}.\arabic{equation}}

\newcommand{\calU}{{\mathcal U}}

\newcommand{\calG}{{\mathcal G}}

\newcommand{\calJ}{{\mathcal J}}
\newcommand{\R}{{\mathbb R}}

\renewcommand{\P}{\mathbb{P}}

\newcommand{\x}{\mathbf{x}}
\newcommand{\z}{\mathbf{z}}
\newcommand{\J}{\widetilde{J}}
\newcommand{\y}{\mathbf{y}}
\renewcommand{\e}{{\mathrm e}}

\newcommand{\pcb}[1]{\textcolor{black}{#1}}
\newcommand{\n}{\mathbf n}

\newcommand{\calT}{{\mathfrak T}}
\newcommand{\calF}{{\mathcal F}}
\renewcommand{\P}{\mathbb P}
\newcommand{\p}{\widetilde{p}}

\newcommand{\f}{\widetilde{f}}

\newcommand{\wrho}{\widetilde{\rho}}
\newcommand{\wphi}{\widetilde{\phi}}

\newcommand{\of}[1]{\textcolor{black}{#1}}

\renewcommand{\S}{\widetilde{S}}
\newcommand{\Q}{\widetilde{Q}}

\begin{document}

 \title[2D target search with boundary-induced resetting ]{2D target search with boundary-induced resetting}

\author{Kevin Chen and Paul C. Bressloff}
\address{Department of Mathematics, Imperial College London, 
London SW7 2AZ, UK}
\ead{p.bressloff@imperial.ac.uk}

\begin{abstract} Stochastic resetting has seen many applications in nonequilibrium statistical physics. The general framework involves a stochastic search process resetting back to its initial state prior to finding the target. Most studies focus on spontaneous resetting whereby resetting events occur at a random sequence of times typically generated by a Poisson process. One of the major consequences of spontaneous resetting is that it yields a finite mean first passage time (MFPT) in unbounded domains, resulting in an optimal resetting rate that minimises the MFPT. An optimal resetting rate can also occur in bounded domains provided that the reset point is not too far from the target. Recently, there has been growing interest in event-based rather than spontaneous resetting, where resetting is triggered when a specific threshold is reached. Identifying the threshold with a physical non-target boundary then leads to so-called boundary-induced resetting. In this paper, we explore the effects of boundary-induced resetting on single-particle diffusive search in a two-dimensional (2D) bounded domain $\Omega$ containing one or more small interior targets. The interior target boundaries are totally absorbing, whereas the exterior boundary $\partial \Omega$ is sticky. That is, whenever the particle reaches a point on $\partial \Omega$ it remains attached for a random waiting time $\tau$ after which it immediately resets to a fixed interior point $\x_0\in \Omega$. Using renewal theory, matched asymptotic analysis and Green's functions
we calculate the unconditional MFPT for absorption by any of the targets and compare this with the corresponding MFPT of a search process on the same domain with a reflecting surface $\partial \Omega$ and no resetting. We show that boundary-induced resetting only reduces the MFPT relative to the no resetting case if the searcher's reset point $\x_0$ is sufficiently close to the targets at $\x_i, i=1,...,N$. The exact condition depends on the geometry of the domain $\Omega$, details of the target configuration such as how close a target is to the sticky boundary, and the mean waiting time $\overline \tau$ at the sticky boundary.

\end{abstract}

\maketitle
\section{Introduction}

Spontaneous stochastic resetting has become a classical paradigm in nonequilibrium statistical physics, see the review \cite{Evans20} and references therein. Broadly speaking, it is based on the notion that a stochastic process resets to its initial state at a random sequence of times $\{T_n,\ n\geq 1\}$. The latter is usually determined by a Poisson process of constant rate $r$ so that the inter-reset times $\tau_n=T_n-T_{n-1}$, $T_0=0$, are independent and exponentially distributed with $\langle \tau_n\rangle =r^{-1}$. One of the main applications of spontaneous resetting is to stochastic search processes where resetting could represent the searcher returning to its home-base prior to finding a target \cite{Pal20}. In the case of Brownian particles diffusing in an unbounded domain, spontaneous resetting renders the mean first passage time (MFPT) finite and there is an optimal resetting rate at which the MFPT is minimised \cite{Evans11a,Evans11b,Evans14}. 
These results apply more broadly to a variety of stochastic search processes, including L\'{e}vy flights \cite{Kus14}, run-and-tumble particles \cite{Evans18}, processes with resetting in a potential landscape \cite{Pal15, Ray20, Roberts24}, partially absorbing targets \cite{Evans13,Schumm21,Bressloff22}, and multiple targets \cite{Chechkin18,Bressloff20,Bressloff22}. On the other hand, further analysis is required to determine when resetting expedites a diffusive search process in a bounded domain, since the MFPT is finite without resetting \cite{Christou15,Durang19,Pal19,Pal19a,Pal22,Mendez22}.

Recently, a number of authors have considered stochastic processes involving so-called {\em event-based resetting}. One version arises in search processes with a partially accessible target, where reaching the boundary of the target is not a sufficient condition for the searcher to access the target (be absorbed) \cite{Bressloff25a,Bressloff25b,Bressloff25c,Bressloff26}. That is, the searcher may stick (adsorb) to the target for some waiting time after which it desorbs and resets to its initial position. Search processes with {\em target-induced resetting} (also known as desorption-induced resetting) can be analysed by extending renewal equations developed for reversible diffusion-controlled reactions \cite{Grebenkov23,Scher23,Scher24}. A second version is {\em threshold resetting} -- a particle resets its state whenever the latter reaches some threshold \cite{DeBruyne20}. This leads to an interesting optimisation problem involving an interplay between the cost of resetting and the benefits of avoiding the threshold. Ref. \cite{Biswas25} explores how threshold resetting affects the fastest FPT of $N$ non-interacting diffusive searchers in the finite interval $[0,L]$ with a target at $x=0$ and the non-target boundary $x=L$ identified with the threshold. Prior to the target being found, all particles reset to a common initial position $x_0$ whenever any particle reaches $x=L$. One finds that for fixed $x_0$ and $N>1$, the fastest FPT has a minimum at an optimal value of $x_0/L$. 

An alternative interpretation of threshold resetting in the interval $[0,L]$ with a target at $x=0$ is that the threshold at $x=L$ is a physical boundary, which leads to the notion of {\em boundary-induced resetting}. It is then natural to compare the MFPT of a single particle with boundary-induced resetting to the case of a reflecting boundary at $x=L$. (In Ref. \cite{Biswas25} the reset-free scenario corresponds to taking the threshold off to infinity, $L\rightarrow \infty$. Although the MFPT is infinite in this limit, the fastest FPT for multiple particles is finite.) As it stands, the 1D case is rather trivial because resetting moves the searcher back towards the target, which will always reduce the MFPT. A more interesting situation holds if a time-cost is accrued whenever the particle hits $x=L$ prior to finding the target. This can be implemented by treating the reactive boundary as sticky. That is, each time the particle hits $x=L$, it adsorbs for some random waiting time $\tau$ after which it desorbs and immediately resets to $x_0$. There then exists a critical mean waiting time $\overline{\tau}_c(x_0/L)$ beyond which resetting is no longer beneficial, that is, the MFPT is no longer reduced. Another nontrivial extension of boundary-induced resetting in 1D is to take the location of the target to be uncertain so that resetting to $x_0$ is not necessarily beneficial \cite{Valla26}. 

In this paper, we explore the effects of boundary-induced resetting on single-particle diffusive search in a two-dimensional (2D) bounded domain $\Omega \subset \R^2$ containing one or more interior targets $\calU$. The interior target boundary $\partial \calU$ is taken to be totally absorbing, whereas the exterior boundary $\partial \Omega$ is taken to be sticky. Moreover, following each desorption event, the particle immediately resets to its initial position $\x_0 \in \Omega \backslash \calU$. We show that, in contrast to the 1D case, boundary-induced resetting without stickiness is not always beneficial when compared to a reflecting boundary $\partial \Omega$. That is, for a fixed domain $\Omega$ and fixed target shape and size, whether or not boundary-induced resetting reduces the MFPT to reach $\partial \calU$ strongly depends on $\x_0$ and the location of the target(s) within $\Omega$. Consequently, the critical mean waiting $\overline{\tau}_c$ of a sticky exterior boundary is a non-trivial function of the underlying geometry.

In order to explore these issues analytically, we will assume that the targets are well separated from each other and the exterior boundary, and are much smaller than the size of the domain. For simplicity, we also assume that the targets are circularly symmetric. The target search problem is then equivalent to a classical 2D narrow capture problem \cite{Coombs09,Ward15,Lindsay16,Bressloff21a}, except that the totally reflecting exterior boundary $\partial \Omega$ is replaced by boundary-induced resetting from a sticky boundary $\partial \Omega$. We proceed along similar lines to our recent work on target-induced resetting \cite{Bressloff25a,Bressloff25b,Bressloff25c} by constructing a first renewal equation that relates the particle propagator with resetting from $\partial \Omega$ to the corresponding propagator for a totally absorbing boundary $\partial \Omega$. We then calculate the Laplace transform of the latter propagator using a combination of matched asymptotic expansions and Green's functions, following along analogous lines to previous studies of 2D narrow capture problems for a totally reflecting boundary $\partial \Omega$\cite{Coombs09,Ward15,Lindsay16,Bressloff21a}. The resulting asymptotic solution is substituted back into the Laplace transformed renewal equation, which is then used to calculate the unconditional MFPT for absorption by one of the targets in the presence of boundary-induced resetting from a sticky boundary $\partial \Omega$. Finally, comparing with the corresponding asymptotic expansion of the MFPT for a reflecting boundary $\partial \Omega$, we derive a condition for boundary-induced resetting to be beneficial in the absence of stickiness and determine the critical mean waiting time when $\partial \Omega$ is sticky. This then allows us to characterise how the effects of boundary-induced resetting depend on the geometry of the underlying search process, as specified by the initial/reset position $\x_0$ of the searcher, the domain $\Omega$, and the centres/radii of the targets. 

The structure of the paper is as follows. We begin by considering the simpler problem of boundary-induced resetting in a finite interval. The corresponding singularly perturbed 2D diffusion problem is formulated in section 3 and the asymptotic analysis is developed in section 4. Finally, we illustrate the theory in section 5 by considering some simple examples.

\setcounter{equation}{0}

\section{Diffusion in a finite interval $[0,L]$}

A classical FPT problem is a Brownian particle with diffusivity $D$ confined to the finite interval $[0,L]$ with an absorbing boundary or target at one end and a reflecting boundary at the other end \cite{Redner01}. For concreteness, suppose that the target is at $x=0$ whereas the reflecting boundary is at $x=L$. The propagator $p_0(x,t|x_0)$ satisfies the forward Kolmogorov or Fokker-Planck equation
\begin{subequations}
\begin{eqnarray}
\label{1Da}
&\frac{\partial p_0(x,t|x_0)}{\partial t}=D\frac{\partial^2p_0(x,t|x_0)}{\partial x^2}, \quad 0<x<L,\\
&p_0(0,t|x_0)=0,\quad  \left .  D\frac{\partial p_0(x,t|x_0)}{\partial x}\right |_{x=L} =0  ,
\label{1Db}
\end{eqnarray}
\end{subequations}
supplemented by the initial condition $p(x,0|x_0)=\delta(x-x_0)$.
The FPT density for being absorbed by (finding) the target at $x=0$ is
\begin{equation}
f_0(x_0,t)=-J_0(0,t|x_0),
\end{equation}
where $J_0(x,t|x_0)=-D\partial_xp_0(x,t|x_0)$ is the rightward probability flux through $x$ at time $t$. The corresponding MFPT is
\begin{equation}
\label{mfptND}
T_0(x_0)=\int_0^{\infty}t f_0(x_0,t)dt=-\partial_s\f_0(x_0,s)|_{s=0},
\end{equation}
with $\f_0(x_0,s)=\int_0^{\infty}\e^{-st}f_0(x_0,t)dt$ the Laplace transformed FPT density. 

Laplace transforming the forward Kolmogorov equation yields the Green's function equation
\begin{subequations}
\begin{eqnarray}
\label{1DLTa}
&D\frac{\partial^2\p_0(x,s|x_0)}{\partial x^2}-s\p_0(x,s|x_0)=-\delta(x-x_0), \quad 0<x<L,\\
& \p_0(0,s|x_0)=0 ,\quad  \left . D\frac{\partial \p_0(x,s|x_0)}{\partial x}\right |_{x=0L} =0 .
\label{1DLTb}
\end{eqnarray}
\end{subequations}
The solution of equations (\ref{1DLTa}) and (\ref{1DLTb}) takes the form
\begin{equation}
\p_0(x,s|x_0)=\left \{ \begin{array}{cc} \frac{\sinh(\sqrt{s/D}x)\cosh(\sqrt{s/D}[L-x_0])}{\sqrt{sD}\cosh(\sqrt{s/D}L)}, & x<x_0\\
\frac{\sinh(\sqrt{s/D}x_0)\cosh(\sqrt{s/D}[L-x])}{\sqrt{sD}\cosh(\sqrt{s/D}L)}, & x>x_0.
\end{array} \right .
\label{GND}
\end{equation}
Moreover, the FPT density for absorption by the target at $x=0$ is
\begin{equation}
\label{f0}
\f_0(x_0,s)=\frac{\cosh(\sqrt{s/D}[L-x_0])}{\cosh(\sqrt{s/D}L)},
\end{equation}
which yields the following well-known expression for the MFPT:
\begin{equation}
\label{MFPT0}
T_0(x_0)=\frac{(2L-x_0)x_0}{2D}.
\end{equation}

 \begin{figure}[t!]
\centering
\includegraphics[width=10cm]{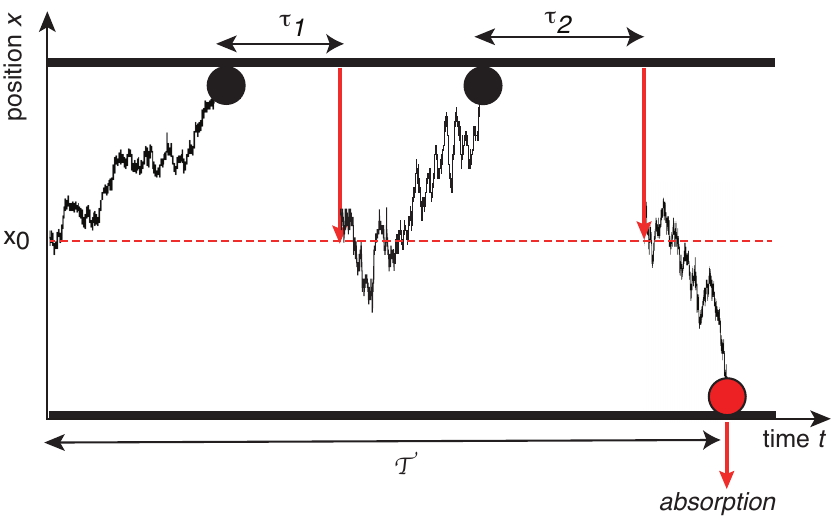} 
\caption{Schematic illustration of a sample path of a Brownian particle in $[0,L]$ with a non-absorbing sticky boundary at $x=L$ and an absorbing target at $x=0$. The particle hits $x=L$ twice prior to finding the target. In each case the particle sticks to the boundary for a random waiting time $\tau_j$, $j=1,2$, after which it desorbs and immediately resets to its initial position $x_0$.}
\label{fig1}
\end{figure}

 Now suppose that whenever the particle hits the boundary at $x=L$ prior to reaching the target at $x=0$, the particle sticks to the right-hand boundary for a random waiting time $\tau$, after which it desorbs and immediately resets to its initial position $x_0$, see Fig. \ref{fig1}. Intuitively speaking, we expect the so-called boundary-induced resetting to reduce the MFPT to find the target, but this is countered by the amount of time spent stuck at $x=L$. Given a general waiting time density $\phi(\tau)$, the propagator $\rho(x,t|x_0)$ for the modified FPT problem satisfies the first renewal equation \cite{Bressloff25a,Bressloff25b}
\begin{eqnarray}
\label{ren1}
 \fl \rho(x,t|x_0)&=p(x,t|x_0)+\int_0^td\tau' \int_{\tau'}^t d\tau\,  \rho(x,t-\tau|x_0)  \phi(\tau-\tau') J(L,\tau'|x_0).
 \end{eqnarray}
The first term on the right-hand side of equation (\ref{ren1}) represents the contribution from all sample paths that have not yet reached the ends $x=0,L$ over the interval $[0,t]$, which is determined by the propagator $p(x,t|x_0)$ of the forward Kolmogorov equation with Dirichlet boundary conditions at both ends:
\begin{subequations}
\begin{eqnarray}
\label{1D2a}
&\frac{\partial p(x,t|x_0)}{\partial t}=D\frac{\partial^2p(x,t|x_0)}{\partial x^2}, \quad 0<x<L,\\
&p(0,t|x_0)=0,\quad   p(L,t|x_0)=0  ,
\label{1D2b}
\end{eqnarray}
\end{subequations}
supplemented by the initial condition $p(x,0|x_0)=\delta(x-x_0)$.
The second term represents all sample paths that reach $x=L$ at least once in the interval $[0,t]$ without being absorbed by the target at $x=0$. The probability of hitting $x=0$ at time $\tau'$, $0<\tau<\tau$, is $J(L,\tau'|x_0)|d\tau'$ where $J(x,t|x_0)=-D\partial_xp(x,t|x_0)$ is the probability flux into the boundary at $x=L$. The particle remains attached to the sticky boundary until desorbing at time $\tau$ with probability $ \phi(\tau-\tau')d\tau$ and then immediately resets to $x_0$. The particle may subsequently return to $x=L$ an arbitrary number of times before reaching $x$ at time $t$. The corresponding FPT density for being absorbed by the target at $x=L$ is
\begin{equation}
\calF(x_0,t)=D\partial_x\rho(0,t|x_0),
\end{equation}
and the MFPT is 
\begin{equation}
\calT(x_0)=-\partial_s\widetilde{\calF}(x_0,s)_{s=0}.
\end{equation}
One subtle point is that the renewal equation (\ref{ren1}) is only valid if $x_0<L$, otherwise the particle never escapes from the sticky boundary. Therefore, we take $x_0\in [0,L-\epsilon]$ with $0<\epsilon \ll  L$ throughout.

The renewal equation can be solved using Laplace transforms and the convolution theorem.
Equation (\ref{ren1}) becomes
\begin{equation}
\wrho(x,s|x_0)=\p(x,s|x_0)+ \wrho(x,s|x_0) \wphi(s) \J(L,s|x_0).
 \end{equation}
 which can be rearranged to give
 \begin{equation}
 \wrho(x,s|x_0)=\frac{\p(x,s|x_0)}{1- \wphi(s) \J(L,s|x_0)}.
 \end{equation}
In addition, Laplace transforming equations (\ref{1D2a}) and (\ref{1D2b}), we find that $\p(x,s|x_0)$ is given by the Dirichlet Green's function
 \begin{equation}
G_D(x,s|x_0)=\left \{ \begin{array}{cc} \frac{\sinh(\sqrt{s/D}x)\sinh(\sqrt{s/D}[L-x_0])}{\sqrt{sD}\sinh(\sqrt{s/D}L)}, & x<x_0\\
\frac{\sinh(\sqrt{s/D}x_0)\sinh(\sqrt{s/D}[L-x])}{\sqrt{sD}\sinh(\sqrt{s/D}L)}, & x>x_0,
\end{array} \right .
\label{GDD}
\end{equation}
and thus
\begin{equation}
\label{J0L}
\fl \J(0,s|x_0)=-\frac{ \sinh(\sqrt{s/D}[L-x_0])}{\sinh(\sqrt{s/D}L)},\quad \J(L,s|x_0)=\frac{ \sinh(\sqrt{s/D}x_0)}{ \sinh(\sqrt{s/D}L)}.
\end{equation}
 It follows that
 \begin{equation}
 \label{calF}
\fl \widetilde{ \calF}(x_0,s)=\frac{|\J(0,s|x_0)|}{1- \wphi(s) \J(L,s|x_0)}=\frac{\sinh(\sqrt{s/D}[L-x_0])}{\sinh(\sqrt{s/D}L)- \wphi(s)\sinh(\sqrt{s/D}x_0)}.
\end{equation}
We will assume that $\phi(\tau)$ has finite moments. Expanding the right-hand side of equation (\ref{calF}) to first order in $s$ with $ \wphi(s)\sim 1-s\overline \tau +O(s^2)$, we find the following expression for the MFPT:
 \begin{equation}
 \calT(x_0)=\frac{L x_0}{2D}+\frac{x_0\overline \tau}{L-x_0}.
 \label{Res1}
 \end{equation}
 
 \begin{figure}[t!]
    \centering
    \includegraphics[width=0.8\linewidth]{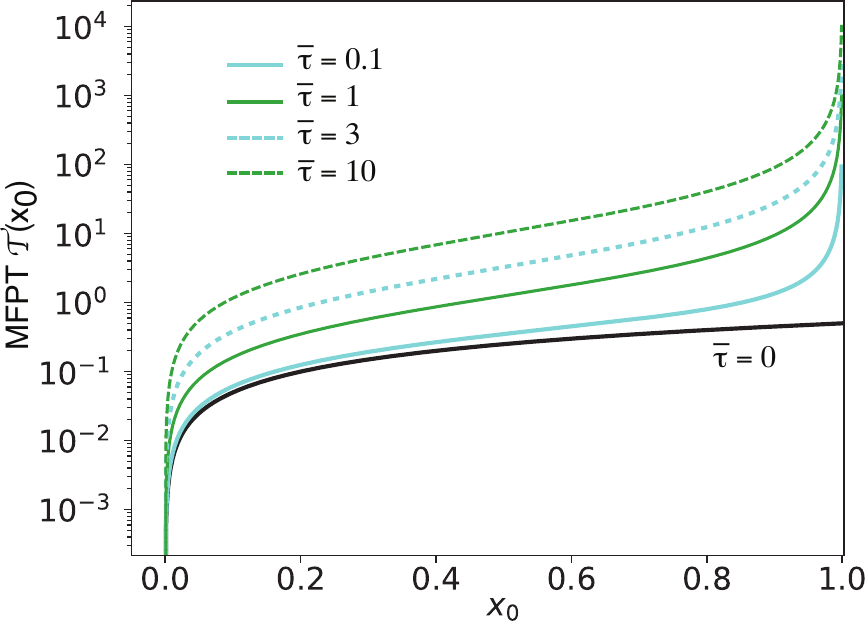}
    \caption{MFPT $\calT(x_0)$ given by equation (\ref{Res1}) plotted on a log-scale with $D, L=1$. The MFPT blows up as $x_0\rightarrow L$ for all $\overline{\tau} > 0$.}
    \label{fig:1D MFPT}
\end{figure}

 \begin{figure}[b!]
    \centering
    \includegraphics[width=0.7\linewidth]{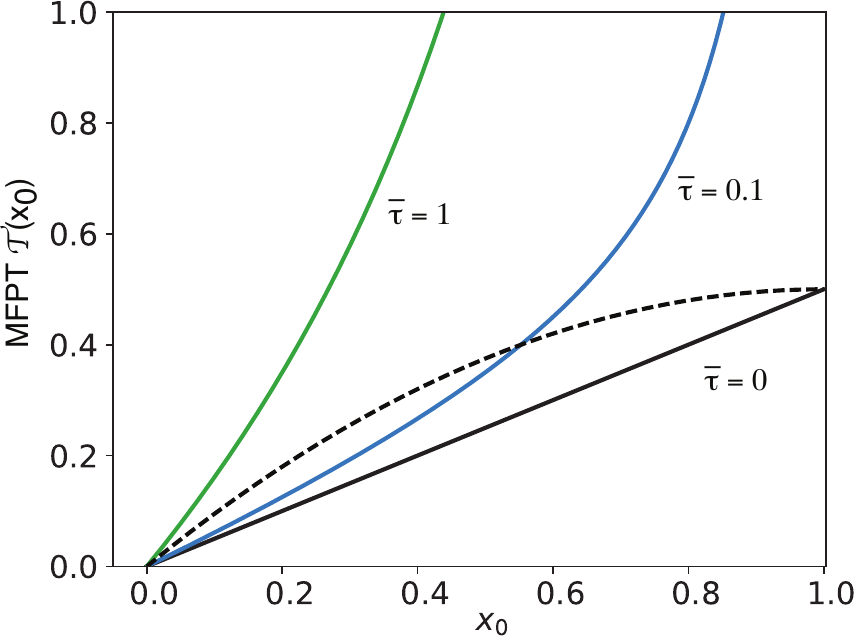}
    \caption{MFPT $\calT(x_0)$ with boundary resetting and a sticky boundary at $x=L$ for various mean waiting times $\overline{\tau}$. The dashed curve shows the classical MFPT $T_0(x_0)$ without resetting and a reflecting boundary at $x=L$. We take $D=L=1$ in all cases so that $\tau_D=1/2$. It can be seen that for $\overline{\tau}<\tau_D$ boundary resetting is advantageous in some part of the domain, but for $\overline{\tau}>\tau_D$, no choice of $x_0$ makes boundary resetting worthwhile.}
    \label{fig:boundary resetting vs resettingless}
\end{figure}

 In  Fig. \ref{fig:1D MFPT} we plot $\calT(x_0)$ as a function of the initial position $x_0$ for various mean waiting times $\overline{\tau}$. When $\overline{\tau}>0$ the MFPT blows up near $x_0=L$. This is as expected since the waiting time on the sticky interface represents a``penalty'' added upon reaching $x=L$, and as $x_0 \rightarrow L$ it is much more likely to hit the sticky interface $x=L$ than the absorbing interface $x=0$. It is also useful to compare the MFPT with boundary resetting, equation (\ref{Res1}) with the classical expression (\ref{MFPT0}). 
 The two MFPTs are equal if
\begin{equation}
     T_0(x_0) = \calT(x_0)\Rightarrow\frac{x_0(2L-x_0)}{2D} = \frac{Lx_0}{2D}+\frac{x_0\overline{\tau}}{(L- x_0)}.
\end{equation}
Since we require $x_0\in [0,L]$, it follows that matching occurs along the critical curve
\begin{equation}
\label{crit}
    \frac{x_0^*}{L} = 1-\sqrt{\frac{\overline{\tau}}{\tau_D}},
\end{equation}
where $\tau_D=L^2/2D$ is the diffusion time-scale. Boundary resetting to $x_0$ is advantageous (reduces the MFPT) provided that $x_0 <x_0^*$. It immediately follows that
boundary resetting is not beneficial for any $x_0$ when $\overline{\tau}\geq \tau_D$. These results are illustrated in Fig. \ref{fig:boundary resetting vs resettingless}. Another way to interpret the critical curve (\ref{crit}) is that for fixed $x_0/L$, there exists a critical mean waiting time 
\begin{equation}
\label{tc1D}
\overline{\tau}_c=\tau_D(1-x_0/L)^2,
\end{equation}
 beyond which boundary-induced resetting is no longer beneficial.

 Finally, we compare the MFPTs after averaging with respect to $x_0\in [0,L-\epsilon]$.
 We find 
 \begin{equation}
\overline T=\frac{1}{L-\epsilon} \int_{0}^{L-\epsilon} T_0(x_0)dx_0=\frac{1}{6D}\bigg (2L^2-\epsilon ( L +\epsilon)\bigg ),
\end{equation}
and
  \begin{equation}
\fl \overline \calT=\frac{1}{L-\epsilon} \int_{0}^{L-\epsilon}\calT(x_0)dx_0=\frac{L(L-\epsilon)}{4D}+\bigg [\frac{L}{L-\epsilon} \ln (L/\epsilon) -1\bigg ]\overline \tau .
\end{equation}
In the latter case, the particle is assumed to reset uniformly to any point in $[ 0,L-\epsilon ]$ after desorbing from $x=L$. To leading order in $\epsilon$, boundary-induced resetting is advantageous provided that $\overline \tau <\overline \tau_c$ with
\begin{equation}
\overline \tau_c\approx \frac{L^2}{12D}\frac{1}{\ln (L/\epsilon)}.
\end{equation}

  \section{Singularly-perturbed diffusion with boundary-induced resetting}
    
    \begin{figure}[b!]
\centering
\includegraphics[width=8cm]{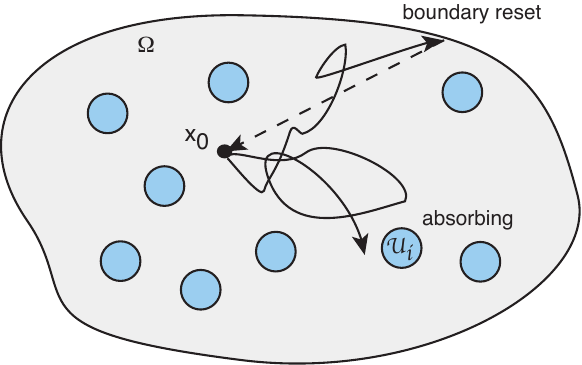} 
\caption{Particle diffusing in a singularly perturbed domain $\Omega\subset \R^2$ containing $N$ totally absorbing interior compartments $\calU_j$ centred at the positions $\x_j$, $j=1,\ldots,N$. The particle resets to its initial position $\x_0$ whenever it desorbs from the exterior boundary $\partial \Omega$.}
\label{figNESS2}
\end{figure}

The effect of boundary resetting on target search in the interval is artificial in the sense that if the particle reaches the non-target end, then resetting away from this boundary is clearly advantageous (at least for a non-sticky boundary). The issue is more complicated for target search in higher spatial dimensions. In particular, consider a 2D simply-connected, bounded domain $\Omega \subset \R^2$, containing a set of $N$ totally absorbing interior targets $\calU_j$, $j=1,\ldots,N$, see Fig. \ref{figNESS2}. Suppose that whenever the particle hits the exterior boundary $\partial \Omega$ it immediately resets to its initial position $\x_0$. We wish to determine the conditions under which boundary resetting reduces the MFPT compared to a totally reflecting exterior boundary. Intuitively speaking, we would expect the answer to depend on the domain $\Omega$, the location and shape of each of the targets and the choice of initial position $x_0$. Another layer of complexity is to determine how 
 the MFPT is affected by stickiness at $\partial \Omega$. In order to explore these issues analytically, we will assume that the targets are well separated and much smaller than the size of the domain. The target search problem is then equivalent to the classical narrow capture problem supplemented by boundary resetting. Hence, we can use matched asymptotic expansions and Green's functions to calculate the MFPT to be absorbed by one of the targets by extending previous studies for a reflecting boundary $\partial \Omega$ \cite{Coombs09,Ward15,Lindsay16,Bressloff21a}.
 
 For concreteness, we take each compartment to be circularly symmetric. Denoting the radius and centre of the $j$-th compartment by $r_j$ and $\x_j$, respectively, we have
\begin{equation}
\calU_j=\{\x\in \Omega,\ |\x-\x_j|< r_j\},\quad \partial \calU_j=\{\x\in \Omega,\ |\x-\x_j|= r_j\}.
\end{equation}
The main characteristic of a singularly perturbed domain is that the compartments $\calU_j$ are small compared to the size of the domain $\Omega$ and are well separated from each other and $\partial \Omega$. More precisely, suppose that the domain $\Omega$ is inscribed by a rectangular area or volume whose smallest dimension is $L$, and introduce the dimensionless parameter
$\epsilon =r_{\max}/L$ where $r_{\max}=\max \{r_j,\, j=1,\ldots,N\}$. We then fix length scales by setting $L=1$ and writing $r_j=\epsilon \ell_j$ with $\ell_j=r_j/r_{\max}$,
and $0<\epsilon \ll 1$. We also assume that $|\x_i-\x_j| =O(1)$ for all $j\neq i$ and $\min_{{\bf s}}\{|\x_j -{\bf s}|,{\bf s} \in \partial \Omega \} =O(1)$, $j=1,\ldots,N$.

Suppose that the exterior boundary is sticky (reversibly adsorbing) with waiting time density $\phi(\tau)$ and whenever the particle desorbs from $\partial \Omega$ ut immediately resets to $\x_0$. Denote the corresponding propagator by $\rho(\x,t|\x_0)$ with $\x,\x_0 \in \Omega\backslash \calU$ and $\calU=\bigcup_{j=1}^N\calU_j$. Analogous to the 1D case, we have the first renewal equation \cite{Bressloff25c}
\begin{equation}
\label{ren2D}
    \rho (\x, t|\x_0) = p(\x,t|\x_0)+\int_0^t d\tau \int_0^\tau \rho(\x, \tau|\x_0)\phi(\tau-\tau') J( \x_0, \tau')d\tau' .
\end{equation}
Here $p(\x, t|\x_0)$ is the propagator of the forward Kolmogorov equation with Dirichlet boundary conditions on $\partial \Omega$ and $\partial \calU$ and no resetting,
\begin{subequations}
    \label{2D p0}
    \begin{eqnarray}
        &\frac{\partial p(\x, t|\x_0)}{\partial t} = D{\bm \nabla}^2p(\x, t|\x_0) \quad \x\in \Omega\backslash \calU,\\
        &p(\x,t |\x_0) = 0, \quad \x\in \partial \Omega \cup \partial \mathcal U ,\\
        &p(\x, 0|\x_0) =\delta(\x-\x_0),
    \end{eqnarray}
\end{subequations}
and $J(\x_0, t)$ is the corresponding probability flux into the exterior boundary,
\begin{equation}
\label{J}
    J(\x_0, t) = -\int_{\partial \Omega} D{\bm \nabla} p(\x, t|\x_0)\cdot \mathbf n(\x) \space d\mathbf x,
\end{equation}
with $\mathbf n(\x)$ the outward pointing normal of the domain $\Omega$ at $\x \in \partial \Omega$. Integrating both sides of equation (\ref{2D p0}a) with respect to $\x\in \Omega\backslash \calU$ and using the divergence theorem implies that
\begin{eqnarray}
\label{Q2D}
\frac{\partial Q(\x_0,t)}{\partial t}=-J(\x_0,t)-\sum_{j=1}^NJ_j(\x_0,t),
\end{eqnarray}
where 
\begin{equation}
Q(\x_0,t)=\int_{\Omega\backslash \calU} p(\x,t|\x_0)d\x
\end{equation}
is the associated survival probability and
\begin{equation}
\label{Jj}
J_j(\x_0,t)= -\int_{\partial \calU_j} D{\bm \nabla} p(\x, t|\x_0)\cdot \mathbf n_j(\x) \space d\mathbf x
\end{equation}
is the flux into the $j$th target. Here $\n_j(\x)$ denotes the unit normal directed towards the interior of $\calU_j$.

We wish to determine the unconditional MFPT $\calT(\x_0)$ that the particle is absorbed by one of the targets in the presence of boundary resetting to $\x_0$. The corresponding FPT is defined according to
\begin{equation}
   {\mathcal T}(\x_0) = \inf \{t\geq0,\mathbf X(t) \in\partial \mathcal U|\mathbf X(0)=\x_0\}.
\end{equation}
Let $\calJ_j(\x_0,t)$ be the total probability flux into the $j$th target,
\begin{equation}
\calJ_j(\x_0,t)=-\int_{\partial \calU_j} D{\bm \nabla} \rho(\x, t|\x_0)\cdot \mathbf n_j(\x) \space d\mathbf x.
\label{calJ}
\end{equation}
The FPT density for ${\mathcal T}(\x_0)$ is then given by
\begin{equation}
\calF(\x_0,t)=\sum_{j=1}^N \calJ_j(\x_0,t),
\label{calF2}
\end{equation}
and
\begin{eqnarray}
\label{MFPT2D}
\fl \calT(\x_0)\equiv \int_0^{\infty} t\calF(\x_0,t)&=-\lim_{s\rightarrow 0} \frac{\partial {\mathcal \calF}(\x_0,s)}{\partial s}
&=-\lim_{s\rightarrow 0} \sum_{j=1}^N \frac{\partial {\mathcal \calJ_j}(\x_0,s)}{\partial s}.
\end{eqnarray}
Hence, it is sufficient to solve the renewal equation (\ref{ren2D}) in Laplace space. The latter takes the form
 \begin{align}
 \label{LTren2D}
 \wrho(\x,s|\x_0)&=\p(\x,s|\x_0) +  \wphi(s) \wrho(\x,s|\x_0) \J(\x_0,s),
 \end{align}
which can be rearranged to give
\begin{equation}
\label{laplace transformed 2D}
 \wrho(\x,s|\x_0) = \frac{\p(\x,s|\x_0)}{1-\wphi(s) \J(\x_0,s)}.
\end{equation}
Moreover, Laplace transforming equations (\ref{2D p0}) gives
\begin{subequations}
\label{p0 transformed 2D}
    \begin{eqnarray}
        & D{\bm \nabla}^2 \p(\x,s|\x_0)-s\p(\x,s|\x_0)=-\delta(\x-\x_0),\quad \x\in \Omega\backslash\calU,\\
        &\p(\x, s|\x_0)=0, \quad \x \in \partial \Omega \cup \partial \calU.    \end{eqnarray}
\end{subequations}
Finally, Laplace transforming equation (\ref{J}) and (\ref{calJ}) and combining with equation (\ref{laplace transformed 2D}) shows that
\begin{equation}
\label{calJs}
 {\mathcal \calJ_j}(\x_0,s) = - \int_{\partial \calU_j} \frac{D{\bm \nabla} \p(\x, s;\x_0)\cdot \mathbf n_j(\x) }{1-\wphi(s) \J(\x_0,s)}\space d\mathbf x   \equiv \frac{\J_j(\x_0,s)}{1-\wphi(s) \J(\x_0,s)} .
\end{equation}

In sections 4 and 5, respectively, we use matched asymptotics to solve the Dirichlet BVP (\ref{p0 transformed 2D}) and then determine $\wrho$ via equation (\ref{laplace transformed 2D}). 
We note that previous studies of the narrow capture problem without boundary resetting have typically taken $\partial \Omega$ to be totally reflecting and the interior targets to be either totally or partially absorbing, That is, equation (\ref{2D p0}) is replaced by the Robin BVP
 \begin{subequations}
 \label{master0}
 \begin{eqnarray}
\fl	&\frac{\partial p(\x,t|\x_0)}{\partial t} = D\nabla^2 p(\x,t|\x_0), \ \x\in \Omega\backslash \calU_a,\quad {\bm \nabla} p \cdot \n=0, \ \x\in \partial \Omega,\\
\fl	&J_j(\y,t|\x_0)\equiv -D{\bm \nabla}p(\y,t|\x_0)\cdot \n_j(\y)=\kappa_0p(\y,t|\x_0) , \quad \y \in \partial \calU_j.
\end{eqnarray}
\end{subequations}
One exception is a study of the 3D narrow capture problem with a Robin boundary condition on $\partial \Omega$ \cite{Coombs15}. However, the focus of this study was to determine how the Robin boundary condition affected the conditional FPT statistics to find a particular target. We have recently incorporated the solution of equation (\ref{master0}a,b) into a renewal equation analogous to (\ref{ren2D}), in which the target surfaces are taken to be sticky rather than $\partial \Omega$ and there is no resetting \cite{Bressloff25c}. When a particle binds to a target (adsorbs),  it remains bound for some waiting time $\tau$, after which it either desorbs or is permanently removed from the system (absorbs). Equation (\ref{ren2D}) becomes
\begin{eqnarray}
 \fl \rho(\x,t|\x_0)&=p(\x,t|\x_0)\\
\fl  &\quad +\int_0^td\tau' \int_{\tau'}^t d\tau\, \sum_{k=1}^N \sigma_{k}\phi_k(\tau-\tau') \bigg [\int_{\partial \calU_k}  \rho(\x,t-\tau|\y)J_k(\y,\tau'|\x_0)d\y \bigg ].\nonumber 
 \end{eqnarray}
 Here $\sigma_k$ is the probability of desorption rather than absorption and $\phi_k$ is the waiting time density at the surface of the $k$th target. The first term on the right-hand side is the contribution from all sample paths that start at $\x_0$ and have not been adsorbed over the interval $[0,t]$. The $k$th contribution on the second line represents all sample paths starting from $\x_0$ that are first adsorbed at some point $\y \in \partial \calU_k$  with flux density $J_k(\y,\tau'|\x_0)$, remain in the bound state until desorbing in the time interval $[\tau,\tau+d\tau]$ with probability $\sigma_k \phi_k(\tau-\tau')d\tau$, after which the particle may bind an arbitrary number of times to various targets before reaching $\x$ at time $t$. In contrast to equation (\ref{ren2D}), the presence of the spatial integrals means that one cannot simply solve the renewal equation using Laplace transforms. However, progress can be made by exploiting the smallness of the targets, see Ref. \cite{Bressloff25c} for further details.
 
\section{Asymptotic analysis in 2D}

\subsection{Matching inner and outer solutions}
We solve equations (\ref{p0 transformed 2D}a,b) in 2D by matching 
`inner' and `outer' expansions in the
limit of small target size $\epsilon\to 0$. In the inner region around the $j$-th target, we introduce the stretched coordinates $\y=(\x-\x_j)/\epsilon$ and set $U_j(\y,s)=\p(\x_j+\epsilon \y,s)$. From equations (\ref{p0 transformed 2D}) we have
\begin{subequations}
\label{2D inner}
\begin{eqnarray}
    &D {\bm \nabla}_\y^2U_j(\y,s)=\epsilon^2 sU_j(\y,s),\quad  |\y|>R_j,\\
    &U_j(\y,s) = 0, \quad |\y|=R_j,
\end{eqnarray}
\end{subequations}
with $R_j=r_j/\epsilon$. Since we will ultimately take the limit $s\rightarrow 0$ in order to calculate the MFPT, we drop the $O(\epsilon^2)$ term. The resulting solution, after using the boundary condition, is of the form
\begin{equation}
    U_j (\y, s) = \nu A_j(\nu,s) \ln(|\y|/R_j),
  \label{2D p0 inner}
\end{equation}
where $\nu = -1/\ln\epsilon$t and the coefficient $A_j (\nu,s) $ will be determined below by matching with the corresponding outer solution. The appearance of $\nu$ as the relevant perturbation parameter is a common feature of 2D singularly perturbed domains \cite{Ward93}. We include the scaling factor $\nu$ for future convenience, anticipating that $A_j(s)=O(1)$.

The outer equation is obtained by shrinking the targets to the single points $\x_j$, $j=1,\ldots,N$, and imposing appropriate singularity conditions:
\begin{subequations}
\label{2D outer}
    \begin{eqnarray}
   \fl     & D{\bm \nabla}^2 \p(\x,s|\x_0)-s\p(\x,s|\x_0)=-\delta(\x-\x_0),\quad \x\in \Omega\backslash  \{\x_1,\ldots,\x_N\},\\
  \fl      &\p(\x, s|\x_0)=0, \quad \x \in \partial \Omega,\\
  \fl  & \p(\x,s|\x_0)\sim \nu A_j(\nu,s)\ln|\x-\x_j|/\epsilon \quad \mbox{as} \ \x\rightarrow \x_j.\end{eqnarray}
\end{subequations}
We solve the outer equation by introducing the Dirichlet Green's function $G(\x, s;{\bm \xi})$ of the modified Helmholtz equation:
\begin{subequations}
\label{2D Greens function}
\begin{eqnarray}
         &D{\bm \nabla}^2 G(\x,s;{\bm \xi})-sG(\x,s;{\bm \xi}) = -\delta(\x-{\bm \xi}),\quad \x, {\bm \xi}\in \Omega,\\
& G(\x,s;{\bm \xi}) = 0, \quad \x \in \partial \Omega.
\end{eqnarray}
\end{subequations}
Note that the Green's function can be decomposed as
\begin{equation}
\label{Greens singular regular 1}
    G(\x, s;{\bm \xi})=-\frac{1}{2\pi D}\ln\left|\x-{\bm \xi}\right| + G_{r\rm eg}(\x, s;{\bm \xi}),
\end{equation}
where $G_{\rm reg}(\x, s;{\bm \xi})$ is the regular part of $G(\x,s; {\bm \xi})$. The solution of equations (\ref{2D outer}) can then be expressed as
\begin{equation}
\label{2D p0 outer}
    \p(\x,s|\x_0)\sim G(\x, s;\x_0)-2\pi \nu D\sum_{j=1}^N A_j(\nu,s)G(\x, s;\x_j).
\end{equation}

The $N$ unknown coefficients $A_j(\nu,s)$ are now obtained by matching the near-field behaviour of the outer solution (\ref{2D p0 outer}) as $\x\rightarrow \x_j$ with the far-field behaviour of the inner solution $\calU_j$ for $j=1,\ldots,N$: This yields the $N$ equations
\begin{eqnarray}
 \fl   G(\x_i, s;\x_0)-2\pi D\nu \sum_{j\neq i} A_jG(\x_i, s;\x_j) - 2\pi D\nu A_iG_{\rm reg}(\x_i, s;\x_i) = A_i-\nu A_i\ln(R_i).\nonumber \\
 \fl
 \label{matching 2D}
\end{eqnarray}
We can write this as a matrix equation:
\begin{equation}
    \l \mathbf I + \nu \bm M(s)\r\mathbf A = \mathbf g(s),
\end{equation}
where $\mathbf A = (A_1, ..., A_N)^\top$, 
$\mathbf g(s) = \l G(\x_1, s;\x_0),,\ldots G(\x_N, s;\x_0)\r^\top$, $\mathbf I$ is the $N$-by-$N$ identity matrix, and $\bm M$ is an $N$-by-$N$ matrix with entries
\begin{subequations}
\label{M entries}
    \begin{eqnarray}
        &\bm M_{ii} (s)= 2\pi DG_{\rm reg}(\x_i, s;\x_i)-\ln R_i, \\
        &\bm M_{ij} (s)= 2\pi D G(\x_i,s; \x_j), & i\neq j.
    \end{eqnarray}
\end{subequations}
If $N=1$, then one solve for $A_1$ explicitly,
\begin{equation}
    A_1(\nu,s) =\frac{G(\x_1,s; \x_0)}{1+\nu\l 2\pi DG_{\rm reg}(\x_1,s; \x_1)-\ln R_1\r}.
\end{equation}
On the other hand, if $N>1$ then the solution requires a matrix inversion:
\begin{equation}
\label{A_i}
    A_i(\nu,s)= \sum_j\l\delta_{ij}+\nu \bm M_{ij}(s)\r^{-1}G(\x_j, s;\x_0),\quad i=1,\ldots,N.
\end{equation}  

\subsection{Expression for the MFPT $\calT(\x_0)$.}

We now use our asymptotic solution for $\p$ to evaluate the MFPT $\calT(\x_0)$ given by equation (\ref{MFPT2D}). We first need to determine the Laplace transformed flux $\calJ_j(\x_0,s)$ into the $j$th target in the presence of boundary resetting, see equation (\ref{calJs}). This, in turn, requires evaluating the flux $\J(\x_0,s)$ into the exterior boundary $\partial \Omega$ in the absence of resetting. The latter is obtained by substituting the outer solution (\ref{2D p0 outer}) into the Laplace transform of equation (\ref{J}), which gives
\begin{equation}
    \J(\x_0, s) \sim-D\int_{\partial \Omega} \bigg [{\bm \nabla} G(\x, s;\x_0)-2\pi \nu D\sum_j A_j(\nu,s){\bm \nabla} G(\x, s;\x_j)\bigg ]\cdot \mathbf n(\x) d\x.
    \label{Jasym}
\end{equation}
A useful alternative expression for $\J(\x_0, s)$ can be obtained by Laplace transforming equation (\ref{Q2D}) and using $Q(\x_0,0)=1$. This gives
\begin{equation}
-\J(\x_0, s)=1-s\Q(\x_0,s)-\sum_{j=1}^N\J_j(\x_0,s).
\end{equation}
The term $\Q(\x_0,s)$ is evaluated by substituting the outer solution (\ref{2D p0 outer}) into the Laplace transform of equation (\ref{Q2D}) so that
\begin{equation}
\label{Q2Ds}
     \Q(\x_0,s) =\int_{\Omega} \bigg [G(\x, s;\x_0)-2\pi \nu D\sum_{j=1}^N A_j(\nu,s) G(\x, s;\x_j)\bigg ]d\x.
\end{equation}
On the other hand, the flux term $\J_j(\x_0,s)$ is obtained by substituting the inner solution (\ref{2D p0 inner}) into the Laplace transform of equation (\ref{Jj}) and evaluating the surface integral
\begin{equation}
\J_j(\x_0,s)=2\pi \nu D A_j(\x_0,s).
\end{equation}
It then follows that
\begin{equation}
\label{Js}
     \J(\x_0,s) = 1-s\int_{\Omega} G(\x, s;\x_0) d\x-2\pi \nu D\sum_{j=1}^N  A_j(\x_0,s)\l1-s\int_{\Omega} G(\x, s;\x_j) d\x\r.
\end{equation}
Finally, Laplace transforming equation (\ref{calJs}) gives
\begin{equation}
\label{calJs2}
 {\mathcal \calJ_j}(\x_0,s) =   \frac{2\pi D\nu A_j(\x_0,s)}{1-\wphi(s) \J(\x_0,s)}.
\end{equation}

Differentiating both sides of equation (\ref{calJs2}) with respect to $s$ yields
\begin{equation}
\label{dcalJ}
\frac{\partial  {\mathcal \calJ_j}(\x_0,s) }{\partial s}= 2\pi D\nu\bigg [  \frac{ \partial_s A_j(\x_0,s)}{1-\wphi(s) \J(\x_0,s)}   + \frac{ A_j(\x_0,s)[\J(\x_0,s)\partial_s\wphi(s)+\wphi(s)\partial_s\J(\x_0,s)]}{[1-\wphi(s) \J(\x_0,s)]^2} \bigg ]  .
\end{equation}
We note that $\wphi(0)=1$ and
\begin{eqnarray}
    \J(\x_0, 0) = 1- 2\pi \nu D\sum_{j=1}^N A_j(\x_0, 0).
\end{eqnarray}
The latter result is obtained by taking the limit $s\rightarrow 0$ in equation (\ref{Js}) and noting that the Dirichlet Green's function is non-singular in this limit. Finally, setting $\partial_s \wphi(0)=-\overline{\tau}$ (assuming that the mean waiting time is finite), we have
\begin{equation}
\label{djds}
    \left.\frac{\partial  {\mathcal \calJ_j}(\x_0,s) }{\partial s}\right| _{s=0} = \frac{\partial _sA_j(\x_0,0)}{\sum_{k=1}^NA_k(\x_0, 0)}+\frac{ A_j(\x_0,0)\l -\overline{\tau}\J(\x_0, 0)+\partial _s \J(\x_0, 0)\r}{2\pi D\nu\l \sum_{k=1}^N A_k(\x_0, 0)\r^2}.
\end{equation}
We thus obtain an asymptotic expression for the MFPT $\calT(\x_0)$ by substituting equation (\ref{djds}) into the right-hand side of equation (\ref{MFPT2D}). One immediate consequence of our analysis is that $\calT(\x_0)=O(1/\nu)$ and this leading order behaviour depends on the mean waiting time $\overline{\tau}$. However, in order to explicitly compute $\calT(\x_0)$ we still need to evaluate $A_j(\x_0, 0)$ and $\partial_s \J(\x_0, 0)$. We will proceed by performing a small-$s$ series expansion of the Dirichlet Green's function.

\subsection{Small-$s$ expansion of the Green's function}

We expand the Dirichlet Greens function (\ref{2D Greens function}) as
\begin{equation}
\label{Gs}
    G(\x, s;{\bm \xi})\sim  G_0(\x;{\bm \xi})+ s G_1(\x; {\bm \xi})+ O(s^2),
\end{equation}
where
\begin{subequations}
\label{Greens expansion 2D}
    \begin{eqnarray}
        &D\nabla^2G_0(\x;{\bm \xi})=-\delta (\x-{\bm \xi}), \quad \x, {\bm \xi}\in \Omega,\\
        &G_0(\x; {\bm \xi}) = 0, \quad \x \in \partial \Omega, \quad {\bm \xi}\in \Omega,
        \end{eqnarray}
        and
        \begin{eqnarray}
        &D\nabla^2G_1(\x;{\bm \xi})=G_0 (\x;{\bm \xi}), \quad \x, {\bm \xi}\in \Omega, \\
        &G_1(\x; {\bm \xi})=0, \quad \x\in\partial \Omega,\quad  {\bm \xi}\in \Omega.
    \end{eqnarray}
\end{subequations}
Here $G_0$ is the Dirichlet Green's function of the Laplacian in $\Omega$ and can be decomposed into a singular and a regular part:
\begin{equation}
G_0(\x;{\bm \xi}) =  -\frac{1}{2\pi D}\ln\left|\x-{\bm \xi}\right| +G_{0,\rm reg}(\x;{\bm \xi}).
\end{equation}
In terms of of $G_0$, the solution of equations (\ref{Greens expansion 2D}c,d) for $G_1(\x, {\bm \xi})$ becomes
\begin{equation}
    G_1(\x; {\bm \xi}) = -\int_\Omega G_0(\x;\mathbf z)G_0(\mathbf z; {\bm \xi})d\mathbf z,
\end{equation}
which is non-singular. Indeed, comparing with equation (\ref{Greens singular regular 1}), we see that
$G_{\rm reg}(\x, s;{\bm \xi})\sim G_{0,\rm reg}(\x; {\bm \xi}) + sG_1(\x, {\bm \xi})+ O(s^2)$. Finally, we make the identifications
\begin{equation}
    G(\x,s=0;{\bm \xi}) = G_0(\x; {\bm \xi}), \quad \frac{\partial G}{\partial s}(\x,s=0;{\bm \xi}) = G_1(\x;{\bm \xi}).
\end{equation}

Plugging the expansion (\ref{Gs}) into the definition of the matrix {\bf M}(s), see equation (\ref{M entries}), gives
    \begin{eqnarray}
    \label{M entries s expanded}
    &\bm M(s) = \bm M^{(0)}+ s\bm \calG + O(s^2),
    \end{eqnarray}
    with
    \begin{subequations}
     \begin{eqnarray}
    \fl  &\bm M_{ii}^{(0)} = 2\pi DG_{0,reg}(\x_i; \x_i)-\ln(R_i),\quad \bm M_{ij}^{(0)} = 2\pi D G_0(\x_i;\x_j),\quad i\neq j,\\
    \fl     &\bm \calG_{ij} = 2\pi D G_1(\x_i; \x_j), \quad \forall i,j.
\end{eqnarray}
\end{subequations}
Keeping $\nu$ fixed, we then have
\begin{eqnarray}
\l \bm I+ \nu \bm M(s)\r^{-1}&= \l \mathbf I+\nu \mathbf M^{(0)}+ \nu s \bm \calG\r^{-1}\nonumber \\
&=\bm \calM -\nu s\bm \calM  \bm\calG\bm \calM + O(s^2),
\end{eqnarray}
where 
\begin{equation}
    \bm \calM  = \l\mathbf I+\nu \mathbf M^{(0)}\r^{-1}.
\end{equation}
Finally, substituting these expressions into equation (\ref{A_i}) yields
\begin{subequations}
\begin{eqnarray}
  \fl   &A_i(\x_0, s=0)= \sum_{j=1}^N \bm\calM_{ij} G_0(\x_j;\x_0),\\
 \fl    &\frac{\partial A_i}{\partial s}(\x_0,s=0) =  \sum_{j=1}^N\left[\bm\calM_{ij} G_1(\x_j; \x_0)-\nu \l\bm\calM\bm\calG\bm\calM\r_{ij} G_0(\x_j;\x_0)\right].
\end{eqnarray}
\end{subequations}
Similarly, differentiating equation (\ref{Js}) with respect to $s$ and taking $s\rightarrow 0$ shows that
\begin{equation}
    \frac{\partial \J}{\partial s}(\x_0,0) = -\int_{\Omega} G_0(\x;\x_0) d\x-2\pi \nu D\sum_{j=1}^N  \left[\partial_sA_j(\x_0,0) - A_j(\x_0, 0)\int_{\Omega} G_0(\x;\x_j) d\x\right].
\end{equation}
Finally, combining with equations (\ref{MFPT2D}) and (\ref{djds}), we have
\begin{eqnarray}
\label{TN}
\fl      \calT(\x_0) &=  \frac{ \overline{\tau}+\int_{\Omega} G_0(\x;\x_0) d\x}{2\pi \nu D \Gamma(\x_0)}-\overline{\tau}- \frac{\sum_{j=1}^NA_j(\x_0, 0)\int_{\Omega} G_0(\x;\x_j) d\x}{\Gamma(\x_0)},
\end{eqnarray}
where
\begin{equation}
\Gamma(\x_0)=\sum_{j=1}^NA_j(\x_0, 0)=\sum_{j,k=1}^N \bm \calM_{jk} G_0(\x_k; \x_0).
\end{equation}

It is important to note that the solution (\ref{TN}) is a non-perturbative function of the small parameter $\nu$, which was obtained by matching the inner and outer solutions using Green's functions along the lines originally developed in Refs. \cite{Ward93}. This effectively sums over the logarithmic terms, which is equivalent to calculating the asymptotic solution for all terms of $O(\nu^k)$ for any $k$. It is $O(1)$ with respect to a corresponding $\epsilon$ expansion. If we now Taylor expand equation (\ref{TN}) in powers of $\nu$, we obrtain the leading $O(1/\nu)$ contribution
\begin{equation}
    \calT(\x_0)\sim \frac{\overline{\tau}+ \int_{\Omega} G_0(\x;\x_0) d\x}{2\pi\nu D\sum_{j=1}^NG_0(\x_j;\x_0)} + O(1),
\end{equation}
where we have used $\bm \calM_{jk}\sim \delta_{jk} + O(\nu)$.

\subsection{MFPT for a reflecting boundary $\partial \Omega$}

As in the 1D case, see section 2, we would like to explore how the MFPT $\calT(\x_0)$ for an active boundary $\partial \Omega$ that includes boundary-induced resetting and stickiness differs from the corresponding MFPT $T(\x_0)$ for a passive reflecting boundary. The latter has been calculated elsewhere \cite{Coombs09,Ward15,Lindsay16,Bressloff21a} so we simply quote the final result here in its Taylor expanded form:
\begin{align}
T(\x_0)&\sim \frac{|\Omega|}{2\pi \nu ND }-\frac{|\Omega|}{N} \bigg [\sum_{j=1}^NG_0(\x_j;\x_0)-\pcb{\frac{1}{N}} \sum_{j,k=1}^N\calG^{(0)}_{kj}\bigg ]+O(\nu).
\label{TP2}
\end{align}
There are then two complementary issues to address. First, in the absence of stickiness ($\overline{\tau}=0$), are there geometric configurations for which boundary-induced resetting is not beneficial, that is, $T(\x_0)<\calT(\x_0)$? Second, in cases where boundary-induced resetting is beneficial without stickiness, what is the analog of the critical curve (\ref{crit}) along which $T(\x_0)=\calT(\x_0)$? If we only keep the leading $O(1/\nu)$ terms in equations (\ref{TN}) and (\ref{TP2}), answering these questions is equivalent to determining the sign of the function
\begin{equation}
\chi(\x_0)=\overline{\tau}+ \int_{\Omega} G_0(\x;\x_0) d\x-\frac{|\Omega|}{N}\sum_{j=1}^NG_0(\x_j;\x_0).
\label{chi}
\end{equation}
That is, boundary-induced resetting is beneficial to leading order if and only if $\chi(\x_0) <0$. We will proceed by considering a few simple target configurations.

\section{Examples}

\subsection{Single target on the unit disk}

\begin{figure}[h!]
    \centering
    \includegraphics[width=0.8\linewidth]{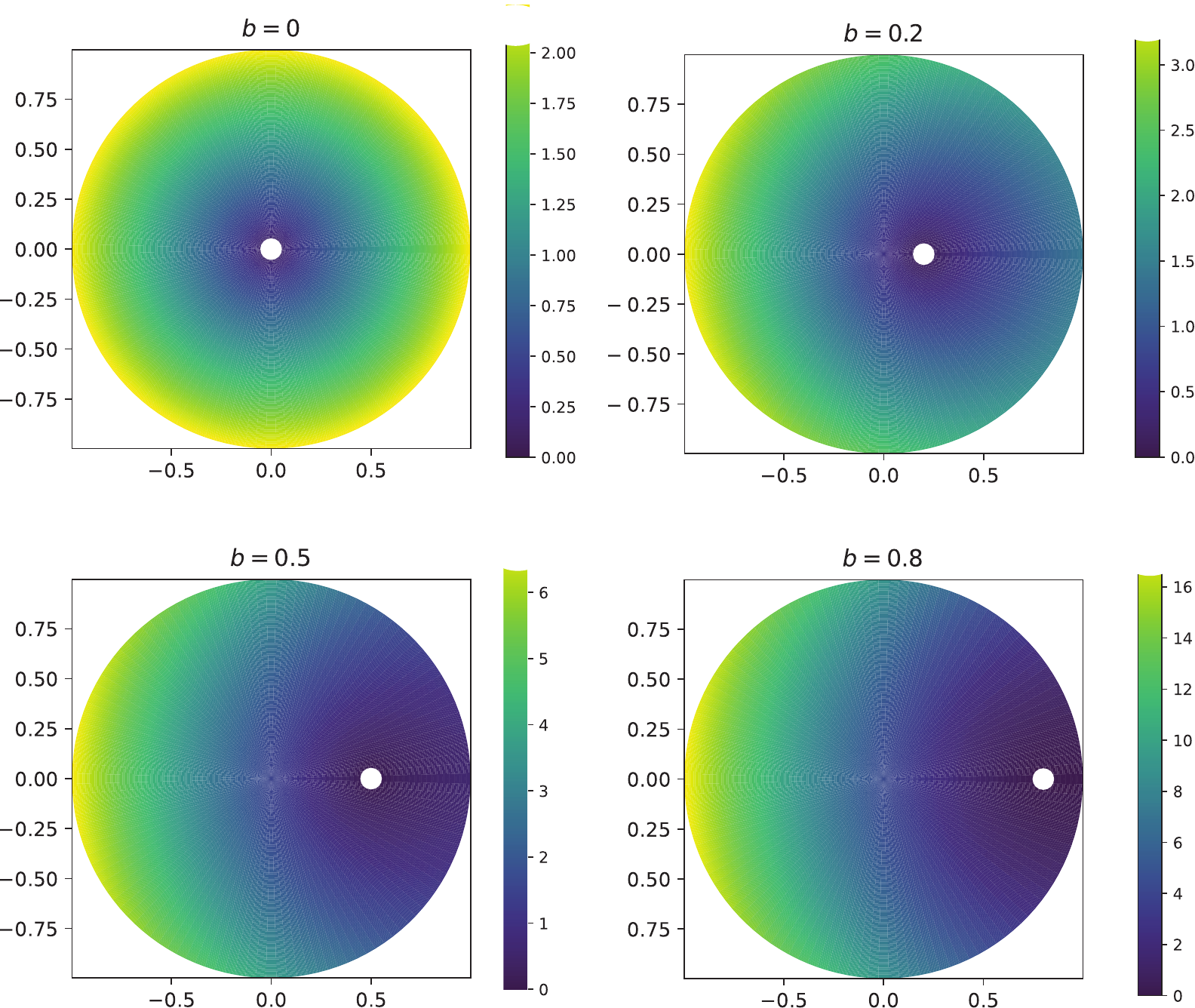}
    \caption{Polar contour plots of the MFPT $\calT(\x_0;\x_1) $ given by equation (\ref{Tone}) as a function of $\x_0=(a\cos \theta,a\sin \theta)$ for different target locations $\x_1=(b,0)$ and $\overline{\tau}=0$. The colouring of each plot is on a log-scale.  Other parameters are $D=1$ and $\nu=0.2$. Top left: $b=0$; Top right: $b=0.3$; Bottom left: $b=0.7$; Bottom right: $b=0.95$. The figure is generated using a discretization in polar coordinates, with a 50-point radial grid taken from $0.001$ to $0.995$, which accounts for singularities and blowups, and a $60$-point angular grid taken from $0$ to $2\pi$.}
    \label{fig5}
\end{figure}

As our first example, consider the unit disk $\Omega$ and a circular target of radius $R_1=r_1/\epsilon = 1$ centred at $\x_1\in \Omega$. The particle begins at $\x_0$ and searches for the target under a Brownian motion. Upon reaching the boundary of the domain $\partial \Omega$, the particle is reset back to $\x_0$ after an exponentially distributed waiting time $\tau$. The Dirichlet Green's function for the diffusion equation on the disk $\Omega$ is well-known:
\begin{subequations}
\begin{eqnarray}
    &G_0(\x; {\bm \xi}) = -\frac{1}{2\pi D}\ln|\x-{\bm \xi}|+\frac{1}{2\pi D}\ln\left|\x|{\bm \xi}|-\frac{{\bm \xi}}{|{\bm \xi}|}\right|,\\
    &G_{0,reg}(\x; {\bm \xi}) = \frac{1}{2\pi D}\ln\left|\x|{\bm \xi}|-\frac{{\bm \xi}}{|{\bm \xi}|}\right|.
\end{eqnarray}
\end{subequations}
Without loss of generality one can exploit the rotational symmetry of the domain and choose $\x_1$ on the positive $x-$axis, such that 
\[\x_0=a\bm e_0, \quad\x_1 = b\bm e_x, \quad a,b\in[0,1),\]
where $\bm e_x$ is the unit vector in the positive $x$ direction, and $\bm e_0$ is a unit vector at an angle $\theta$ with respect to $\bm e_x$:
\[\bm e_x\cdot \bm e_0 = \cos\theta.\]
We then have
\begin{eqnarray}
\fl &G_0(\x_1;\x_0) = -\frac{1}{4\pi D}\ln\l a^2+b^2-2ab\cos\theta\r+\frac{1}{4\pi D}\ln\l |a^2b^2+1-2ab\cos\theta|\r.
\end{eqnarray}
Using equation (\ref{TN}), the MFPT for a single target is
\begin{eqnarray}
     \calT(\x_0;\x_1) &=  \frac{\overline{\tau}+\int_{\Omega} G_0(\x;\x_0) d\x}{2\pi \nu D  \calM_{11} G_0(\x_1; \x_0)}-\overline{\tau}- \int_{\Omega} G_0(\x;\x_1) d\x,
     \label{Tone}
\end{eqnarray}
with
\begin{equation}
\calM _{11}\sim \frac{1}{1+2\pi \nu D G_{0,\rm reg}(\x_1; \x_1)} = \frac{1}{1+\nu \ln[1-b^2]}.
\end{equation}
The dependence of the MFPT on the target location $\x_1$ has been made explicit.
Note that $\calM _{11}$ blows up as $\nu \ln\l1-b^2\r \rightarrow -1$ from above. However, recall that one of the necessary conditions for the validity of the asymptotic analysis (see second paragraph of section 3) is that each target has to be well-separated from the boundary $\partial \Omega$. Imposing the condition $-\nu \ln\l1-b^2\r <1$ and exponentiating with $\e^{-1/\nu}=\epsilon$, we obtain the equivalent condition $b^2+\epsilon -1< 0$, which holds provided that $b< 1-\epsilon/2$. In Fig. \ref{fig5} we show polar contour plots of the MFPT $\calT(\x_0;\x_1) $ as a function of $\x_0=(a\cos \theta,a\sin \theta)$ for different target location $\x_1=(b, 0)$ and $\overline{\tau}=0$ (non-sticky boundary $\partial \Omega$). In order to avoid singularities, we have picked values of $b$ such that $0<b<1-\epsilon$. 

\begin{figure}
    \centering
    \includegraphics[width=\linewidth]{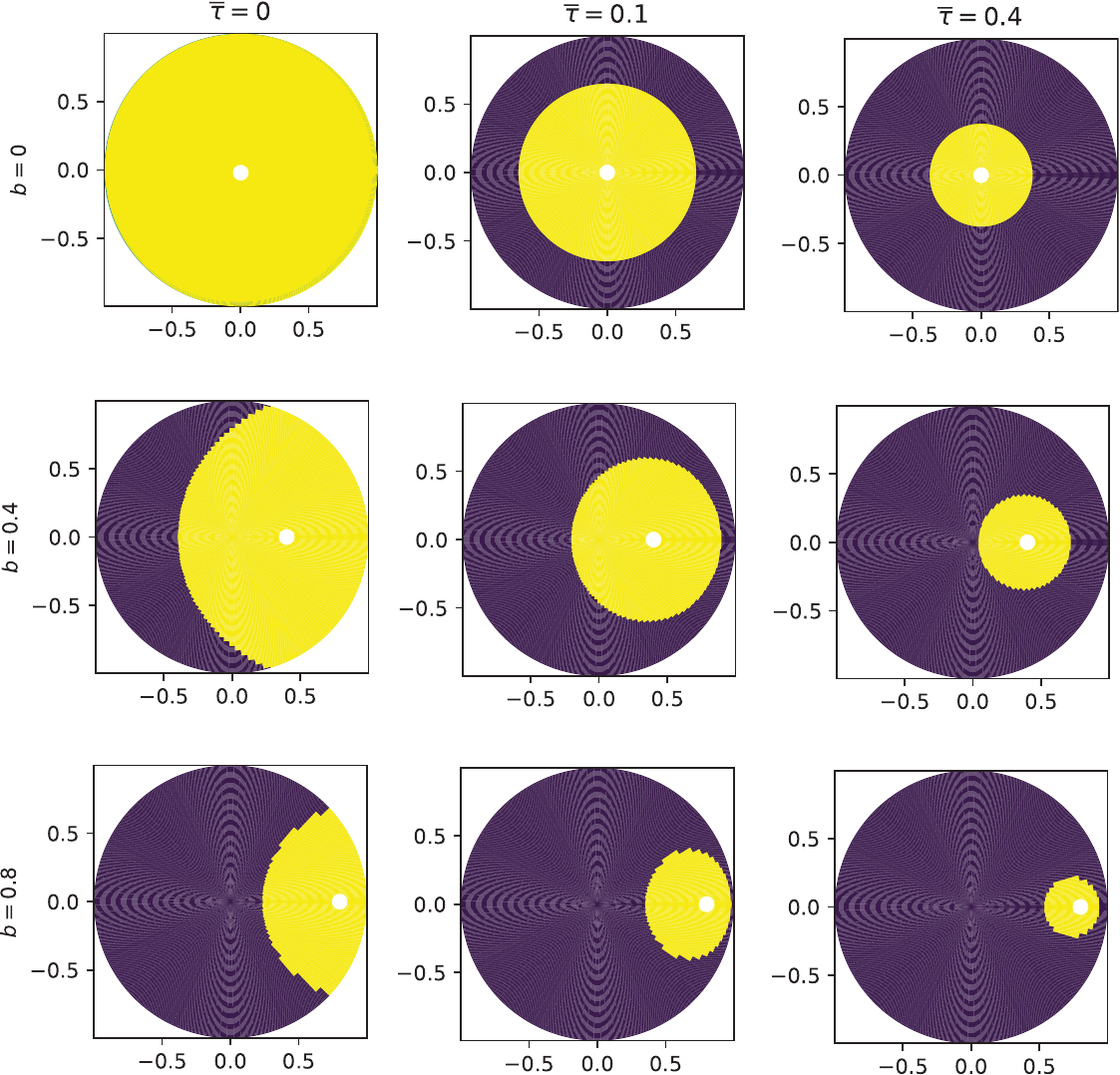} 
    \caption{Comparison of the $O(1/\nu)$ contributions to the MFPT with and without boundary-induced resetting, respectively. Polar contour plot of $\mbox{sign}(\chi(\x_0;\x_1))$, see equation (\ref{2Dcriterion}), as a function of  $\x_0=(a\cos \theta,a\sin \theta)$ for different target locations $\x_1=(b,0)$ and mean waiting times $\overline{\tau}$. The plots are generated in discretised polar coordinates, with a 150-point radial grid from $0.001$ to $0.995$ and a 120-point angular grid from $0$ to $2\pi$, and locations where $\x_0, \x_1$ are no longer well-separated are excluded. Boundary-induced resetting is advantageous (disadvantageous) in the light (dark) shaded regions.}
    \label{fig6}
\end{figure}

In the case of a single target in the unit disk, we have $|\Omega|=\pi, N=1$ and equation (\ref{chi}) becomes
\begin{equation}
\label{2Dcriterion}
  \chi(\x_0;\x_1)=\overline{\tau}+ \int_{\Omega} G_0(\x;\x_0) d\x-\pi G_0(\x_1; \x_0).
\end{equation}
It immediately follows that if $\overline{\tau}=0$, then boundary-induced resetting is beneficial if and only if 
\begin{equation}
     \int_{\Omega} G_0(\x;\x_0) d\x<\pi G_0(\x_1; \x_0).
\end{equation}
In the limit $|\x_1|\rightarrow 1$, $G_0(\x_1, \x_0)\rightarrow0$ while the integral $\int_{\Omega} G_0(\x,\x_0) d\x$ remains positive and fixed, meaning that for a target centre $\x_1$ fclose to the exterior boundary, boundary-induced resetting is only advantageous if $\x_0$ is sufficiently close to $\x_1$. This is illustrated in Fig. \ref{fig6}, which shows polar contour plots of $\mbox{sign}(\chi(\x_0;\x_1)) $ as a function of $\x_0=(a\cos \theta,a\sin \theta)$ for different target locations $\x_1=(b,0)$ and $\overline{\tau}$. Note that for $b>0$, there exist values of $\x_0$ where boundary-induced resetting is disadvantageous. This contrasts with the 1D scenario, where boundary-induced resetting is always beneficial, since resetting always brings the particle closer to the target. In 2D, one sees that boundary-induced resetting is only advantageous if the initial position is "close enough" to the target, where the definition of "close enough" depends on the location of the target and the mean waiting time $\overline \tau$. For some values of $\x_1$, such as if $|\x_1| = b=0.8$, one sees that even an initial position from the centre of the circular domain $\x_0=0$ won't benefit from boundary-induced resetting, and the latter is advantageous only if the choice of $\x_0$ actively favors the location of the target $\x_1$.
On the other hand, the $b=0$ cases are analogous to a 1D model due to radial symmetry. In particular, boundary-induced resetting is advantageous everywhere when $\overline \tau=0$.

\begin{figure}[t!]
    \centering
    \includegraphics[width=0.6\linewidth]{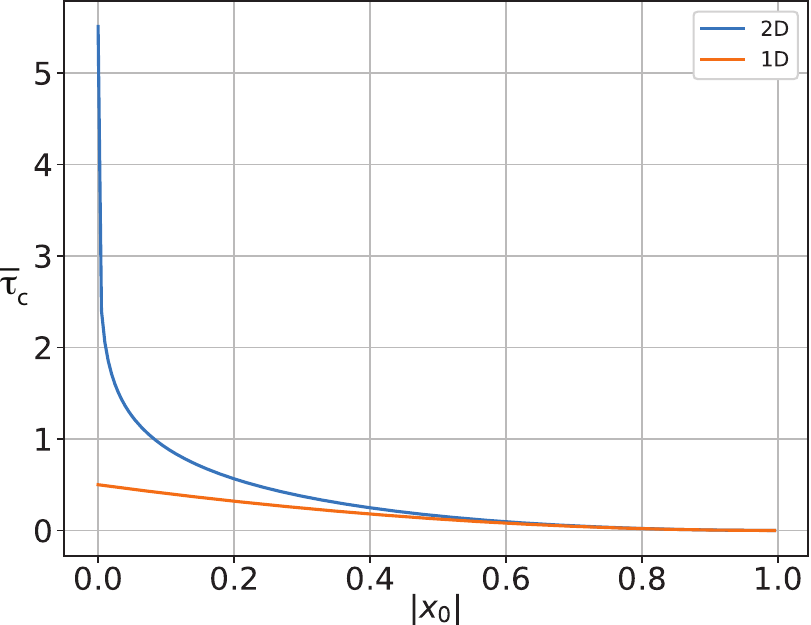} 
    \caption{Critical curve $\overline{\tau}_c(\x_0;\x_1)$ for $\x_1=(0,0)$ plotted as a function of $|\x_0|$, compared between 1D and 2D cases.}
    \label{fig7}
\end{figure}

  \begin{figure}[b!]
    \centering
    \includegraphics[width=0.9\linewidth]{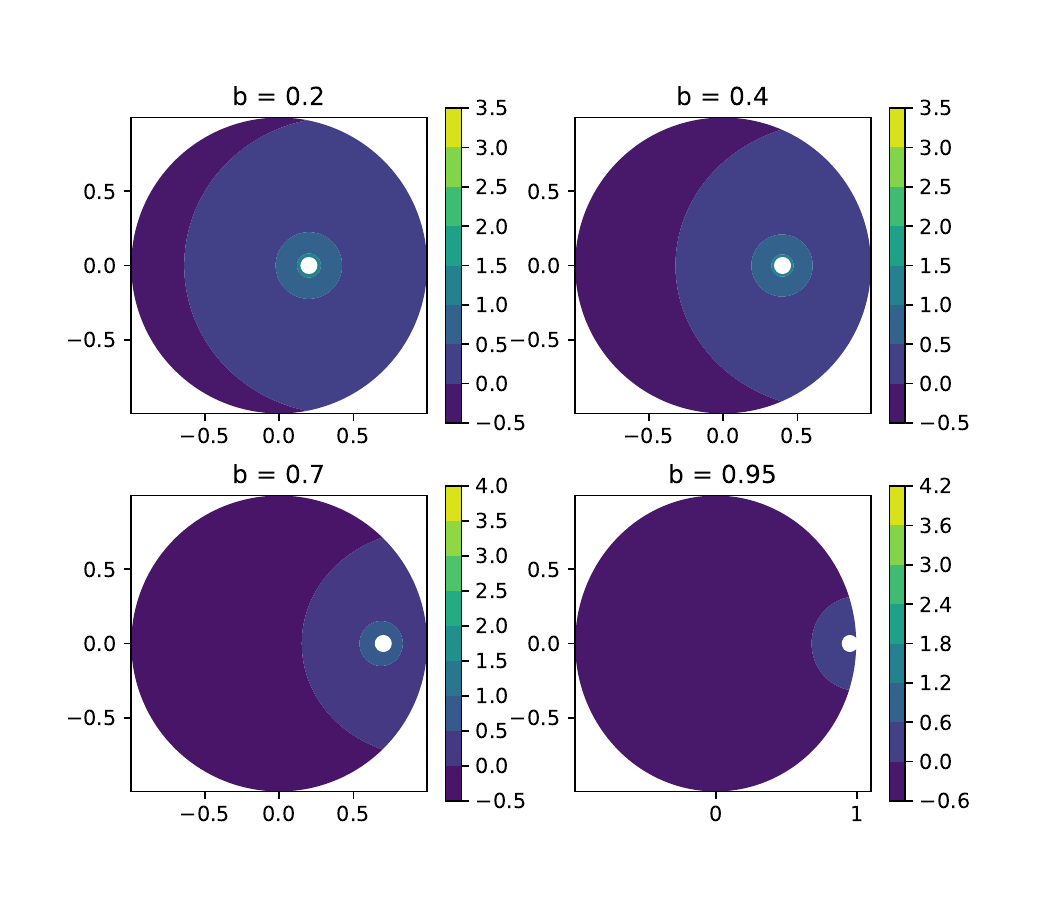} 
    \caption{Polar contour plots of $\overline{\tau}_c(\x_0;\x_1)$ as a function of $\x_0=(a\cos \theta,a\sin \theta)$ for different target locations $\x_1=(b,0)$ with $a>0$.The discretisation and exclusion schemes are the same as Fig. \ref{fig6}.}
    \label{fig8}
    \end{figure}

Similar to equation (\ref{tc1D}), the critical waiting time $\overline{\tau}_c(\x_0;\x_1)$ is defined by the condition $\chi(\x_0;\x_1)=0$, so that to $O(1/\nu)$ we have
\begin{equation}
\overline{\tau}_c(\x_0;\x_1)=\pi G_0(\x_1; \x_0)-\int_{\Omega} G_0(\x;\x_0) d\x.
\end{equation}
In Fig. \ref{fig7} we plot $\overline{\tau}_c(\x_0;\x_1)$ for $\x_1=0$ as a function of $|\x_0|$ alone, since we can exploit radial symmetry. On the same figure we plot the one-dimensional critical curve $\overline{\tau}_c = \overline{\tau}_c(x_0)$ from (\ref{tc1D}). One sees that in general, the 2D problem adopts a larger $\overline \tau_c$ than the 1D counterpart, which is due to simply having extra dimensions for the searcher to travel in. On the other hand, one notes that in both problems $\overline \tau_c>0$ for all values of $|\x_0|$. This is consistent with Fig. \ref{fig6} where if  $\overline \tau = 0$ and the target is located at the centre of the circular domain, then boundary-induced resetting will always bring the searcher closer to the target as long as $|\x_0|<1$. It is clear that this does not extend to cases where the target is not located at the centre of the domain: In Fig. \ref{fig8} we show corresponding polar contour plots of the 2D $\overline{\tau}_c(\x_0;\x_1)$ for fixed $|\x_1|>0$, where we observe negative values of $\overline\tau_c$ in some regions, corresponding to the regions in \ref{fig6} where boundary-induced resetting is disadvantageous. For the regions where $\overline \tau_c>0$, one sees that a mean waiting time $\overline \tau = O(1)$ will be large enough for boundary-induced resetting to be disadvantageous. This particular scaling is again due to the small target size: Since one expects multiple arrivals at the sticky boundary before the particle finds the target, the cumulative waiting time blows up as $\epsilon \rightarrow 0$ which slows down the boundary resetting process.

\subsection{Symmetric pair of targets on the unit disk}

\begin{figure}[b!]
    \centering
    \includegraphics[width=\linewidth]{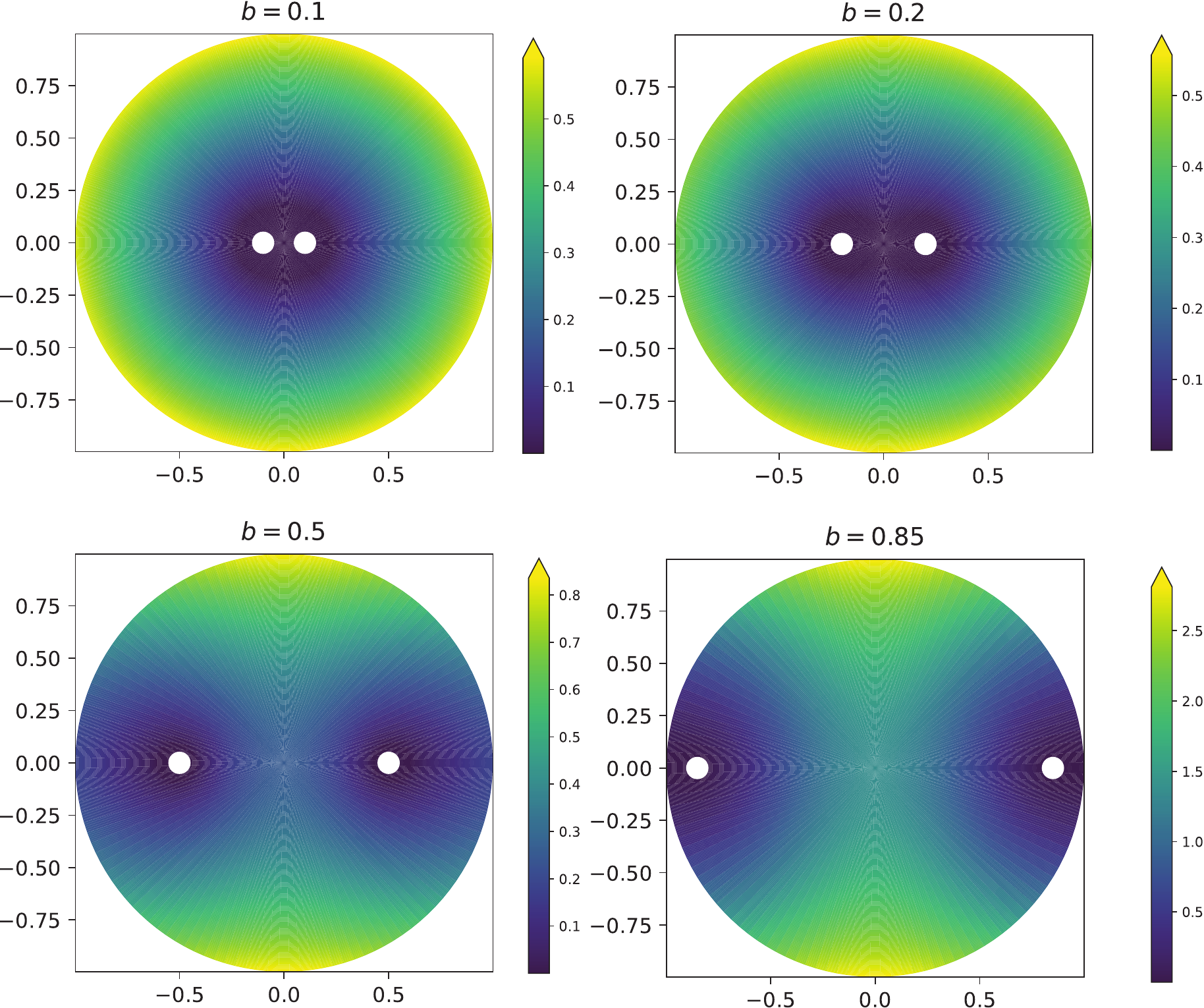}
    \caption{Polar contour plots of the MFPT $\calT(\x_0;\x_1) $ given by equation (\ref{MFPT Dumbell}) as a function of $\x_0=(a\cos \theta,a\sin \theta)$ for a pair of identical targets at $\x_1 = b\bm e_x, \quad \x_2 =-\x_1 -b\bm e_x$ (symmetric dumbbell configuration). The colouring of each plot is on a log-scale.  Other parameters are $D=1$ and $\nu=0.2$. Top left: $b=0.1$; Top right: $b=0.2$; Bottom left: $b=0.$; Bottom right: $b=0.85$. The discretization scheme used is the same as Fig.  \ref{fig5}.}
    \label{fig9}
\end{figure}

\begin{figure}[t!]
    \centering
    \includegraphics[width=\linewidth]{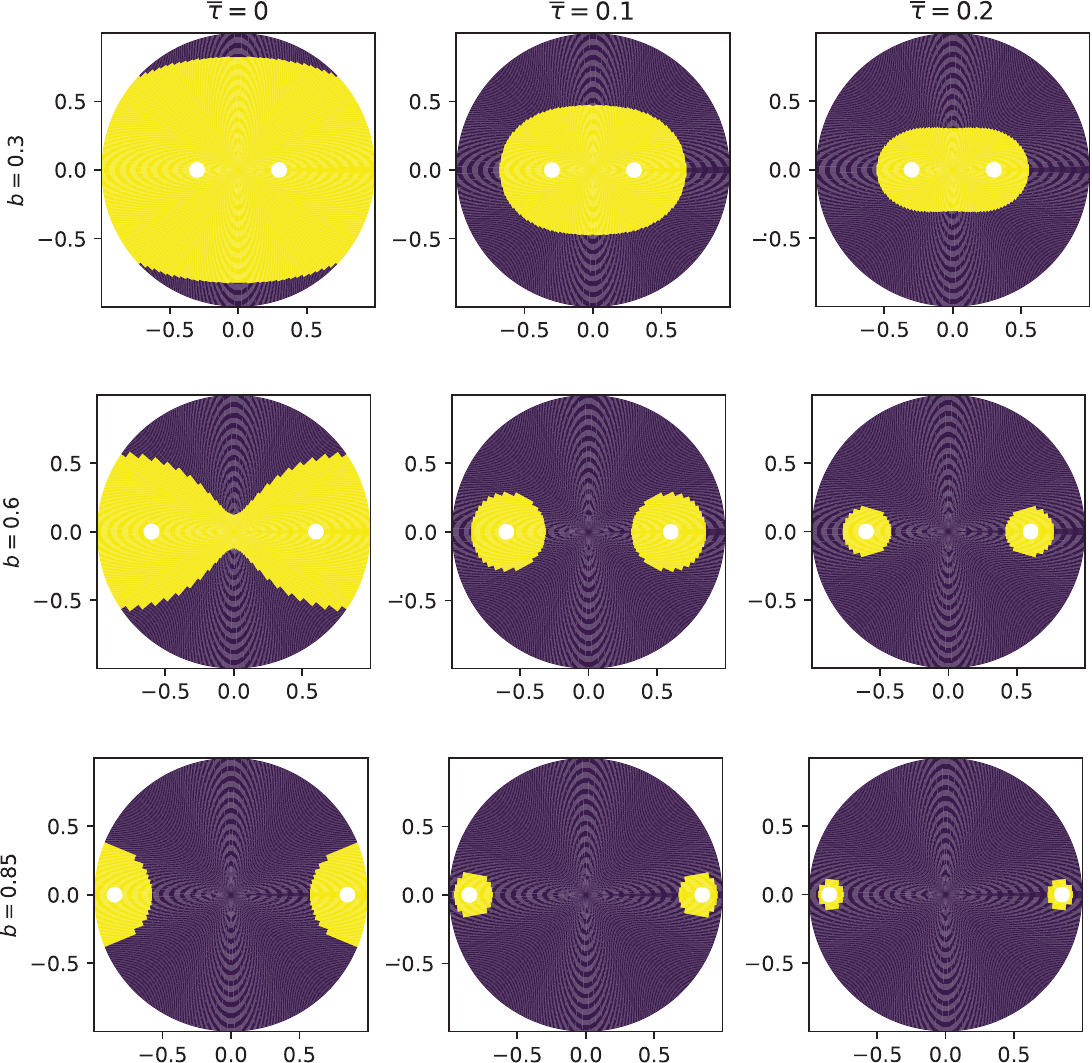}
    \caption{Comparison of the $O(1/\nu)$ contributions to the MFPT with and without boundary-induced resetting, respectively. Polar contour plot of $\mbox{sign}(\chi(\x_0;\x_1))$  as a function of  $\x_0=(a\cos \theta,a\sin \theta)$ for the symmetric dumbbell configuration and mean waiting times $\overline{\tau}$, see Eq. (\ref{dumbell criterion}). Boundary-induced resetting is advantageous (disadvantageous) in the light (dark) shaded regions.
Other parameters are as in Fig. \ref{fig9}}
    \label{fig10}
\end{figure}

We now consider two identical targets in the unit disk $\Omega$ with $R_1 = R_2=1$, centred at symmetric locations about the centre of the disk, such that $\x_2 = -\x_1 $, where $\x_1, \x_2\in\Omega$. Analogous to the previous example, we define
\[\x_0=a\bm e_0, \quad\x_1 = b\bm e_x, \quad \x_2 = -\x_1=-b\bm e_x, \quad a,b\in[0,1], \]
with $\bm e_x$ the unit vector in the positive $x$ direction, and $\bm e_0$ some unit vector at an angle $\theta$ with $\bm e_x$, such that
\[\bm e_0 \cdot \bm e_x=\cos(\theta).\]
One notices from the radial symmetry of the domain $\Omega$ that
\[I(\x_1):=\int_\Omega G_0(\x, \x_1)d\x =\int_\Omega G_0(\x, -\x_1)d\x ,  \]
which implies
\begin{equation}
     \sum_{j=1}^NA_j(\x_0, 0)\int_{\Omega} G_0(\x;\x_j)d\x= I(\x_1)\Gamma(\x_0,\x_1),
\end{equation}
where we have made the dependence on the target coordinate $\x_1$ explicit.
The MFPT (\ref{TN}) therefore simplifies to
\begin{equation}
\label{MFPT Dumbell}
\calT(\x_0,\x_1) =  \frac{ \overline{\tau}+\int_{\Omega} G_0(\x;\x_0) d\x}{2\pi \nu D \Gamma(\x_0,\x_1)}-\overline{\tau}- I(\x_1).
\end{equation}
It remains to calculate the matrix $\bm \calM$, and the value of $\Gamma(\x_0,\x_1)$. For this configuration, we have that
\begin{align}
    \bm \calM  &= \l\mathbf I+\nu \mathbf M^{(0)}\r^{-1}\nonumber \\
        &=\begin{pmatrix}
            1+  \nu\ln(1-b^2)&  -\nu \ln(2b)+\nu \ln(1+b^2)\\
           -\nu \ln(2b)+\nu \ln(1+b^2) &1+  \nu\ln(1-b^2)\\
        \end{pmatrix}^{-1}\nonumber \\
        &=\frac{1}{\Delta}\begin{pmatrix}
            1+  \nu\ln(1-b^2)&  \nu \ln(2b)-\nu \ln(1+b^2)\\
           \nu \ln(2b)-\nu \ln(1+b^2) &1+  \nu\ln(1-b^2)\\
        \end{pmatrix},
    \end{align}
where $\Delta$ is the determinant of the matrix before inversion:
\begin{eqnarray}
    \Delta = (1+\nu\ln(1-b^2))^2-\nu^2\l\ln(2b)-\ln(1+b^2)\r^2.
\end{eqnarray}
We can therefore calculate the quantities used in (\ref{TN}):
\begin{eqnarray}
\fl \Gamma(\x_0,\x_1) &= \frac{2}{\Delta}\l1+\nu \ln(1-b^2) +\nu\l\ln(2b)-\ln(1+b^2)\r\r\l G_0(\x_1, \x_0)+ G_0(-\x_1, \x_0)\r\nonumber \\
 \fl   & = \frac{2\l G_0(\x_1, \x_0)+ G_0(-\x_1, \x_0)\r}{1+\nu\ln(1-b^2)-\nu(\ln(2b)-\ln(1+b^2))},
\end{eqnarray}
where
\numparts
\begin{eqnarray}
\fl &G_0(\x_1;\x_0) = -\frac{1}{4\pi D}\ln\l a^2+b^2-2ab\cos\theta\r+\frac{1}{4\pi D}\ln\l a^2b^2+1-2ab\cos\theta\r,\\
\fl &G_0(-\x_1;\x_0) = -\frac{1}{4\pi D}\ln\l a^2+b^2+2ab\cos\theta\r+\frac{1}{4\pi D}\ln\l a^2b^2+1+2ab\cos\theta\r.
\end{eqnarray}
\endnumparts

In Fig. \ref{fig9} we plot the MFPT of the dumbbell target problem with various initial starting position $\x_0$, analogous to Fig. \ref{fig5}. Note that having an extra target drastically reduces the MFPT compared to the single-target case. Equation (\ref{TN} ) suggests that the leading order contribution to the MFPT is approximately halved with 2 targets.
 Assuming that $\nu$ is small enough such that $O(1)$ terms may be ignored, we can set $N=2$ to deduce that boundary-induced resetting is only beneficial if
\begin{equation}
\label{dumbell criterion}
    \chi(\x_0,\x_1) \equiv \overline \tau+ \int_{\Omega} G_0(\x,\x_0) d\x-\frac{\pi}{2} \bigg [G_0(\x_1 \x_0)+G_0(-\x_1 \x_0)\bigg ]<0.
\end{equation}
Fig. \ref{fig10} depicts the comparison between the two schemes. Similar to the single-target case, boundary-induced resetting is only advantageous if at least one of the targets is sufficiently close to the initial position. One also notice a form of separation analogous to that of droplet formation. The separation is observed in the intermediate transition as $\overline \tau$ grows, between the large $\overline \tau$ behaviour where boundary resetting is only faster if the particle starts sufficiently close to one of the target, and the low $\overline \tau$ behaviour where the clusters are allowed to be further away for boundary resetting to still be advantageous.

\section{Discussion}

In this paper we studied a stochastic search process with boundary-induced resetting on a 2D domain $\Omega\subset \R^2$ with a sticky exterior surface $\partial \Omega$ and one or more small totally absorbing interior targets. We calculated the unconditional MFPT for absorption by any of the targets and compared this with the corresponding MFPT of a search process on the same domain with a reflecting surface $\partial \Omega$ and no resetting. Our results imply that boundary-induced resetting only reduces the MFPT relative to the no resetting case if the searcher's reset point $\x_0$ is "sufficiently close" to the targets at $\x_i, i=1,...,N$. The exact condition depends on the geometry of the domain $\Omega$, details of the target configuration such as how close a target is to the sticky boundary, and the size of the mean waiting time $\overline \tau$ at the sticky boundary. 

One natural extension of the current work is to consider the analogous 3D narrow capture problem with boundary-induced resetting. Although we would expect analogous qualitative results to hold with regards the effectiveness of boundary-induced resetting in reducing the MFPT, the details of the matched asymptotic analysis will differ significantly. This reflects the fact that the Green's function of the Laplacian has a different singularity structure in 2D and 3D: 
\begin{align*}
&G(\x,\x_0)\rightarrow -\frac{1}{2\pi D}\ln|\x-\x_0| \mbox{ in 2D },  \\ &G(\x,\x_0)\rightarrow \frac{1}{4\pi D|\x-\x_0|} \mbox{ in 3D}
\end{align*}
as $|\x-\x_0|\rightarrow 0$.
Hence, an asymptotic expansion of the solution to a narrow capture problem in 3D is in powers of $\epsilon$ rather than $\nu=-1/\ln \epsilon$ \cite{Cheviakov11,Chevalier11,Coombs15,Bressloff21b}.
Another possible extension of the model is to consider more general sticky boundary conditions on $\partial \Omega$. First, we could take a non-spatially-uniform waiting time distribution along the exterior surface $\Omega$. This would require keeping track of the position $\z\in \partial \Omega$ where the particle hits the boundary in the renewal equation (\ref{ren2D}) and considering a spatially heterogeneous waiting time density $\phi(\tau, \z)$. Second, we could modify the mechanism of how the particle attaches to the wall, that is, how it enters the sticky state. This would involve replacing the Dirichlet boundary condition in equation (\ref{2D p0}b) by a partially absorbing Robin boundary condition, say \cite{Bressloff25a,Bressloff25b,Bressloff25c}.

\end{document}